\documentclass[twocolumn,nofootinbib,aps,prd]{revtex4-2}

\usepackage{amsfonts,amsmath,amssymb,mathrsfs}
\usepackage{mathptmx,mathtools,latexsym}
\usepackage{xcolor}
\usepackage[colorlinks=true,linkcolor=blue,citecolor=blue,urlcolor=blue]{hyperref}
\usepackage{orcidlink}

\begin{document}

\title{Two-horizon sector thermodynamics and NUT bi-hair in Kerr-Newman-NUT spacetimes}

\author{Di Wu\orcidlink{0000-0002-2509-6729}$^{1,2}$}
\email{Contact author: wdcwnu@163.com, d277wu@uwaterloo.ca}
%https://orcid.org/0000-0002-2509-6729

\author{Robert B. Mann\orcidlink{0000-0002-5859-2227}$^2$}
\email{Contact author: rbmann@uwaterloo.ca}
%https://orcid.org/0000-0002-5859-2227

\author{Shuang-Qing Wu\orcidlink{0000-0001-7936-7195}$^1$}
\email{Contact author: sqwu@cwnu.edu.cn}
%https://orcid.org/0000-0001-7936-7195

\affiliation{$^1$School of Physics and Astronomy, China West Normal University,
Nanchong, Sichuan 637002, People's Republic of China \\
$^2$Department of Physics and Astronomy, University of Waterloo, Waterloo, Ontario N2L 3G1, Canada}

\date{\today}

\begin{abstract}
First law consistency alone does not determine the black hole state space of
NUT-charged spacetimes, because several inequivalent macroscopic formulations can yield
consistent thermodynamics. We develop a two-horizon sector diagnostic for the
electrically charged Kerr-Newman-NUT family by organizing the inner- and outer-horizon
entropies and inverse temperatures into sum ($\Sigma$) and difference ($\Delta$) sectors.
The charged rotating geometry gives a more stringent test of the
uncharged Taub-NUT sector result, since the mass $M$, ordinary electric charge $Q$,
ordinary angular momentum $J$, NUT charge $N$, and the NUT secondary hair $J_N=MN$
all enter the same
two-horizon thermodynamics. For the physically anchored minimal homogeneous state spaces
defined and tested here,
we find a clear sector separation: the
$\Sigma$ sector closes with the charge-like variables $Q$ and $N$, whereas the
$\Delta$ sector additionally requires the rotation-like variables $J$ and $J_N$.
Natural reductions that eliminate either $J_N$ or $N$ fail to reproduce the first law
coefficient fixed by the sector temperature, and their associated Bekenstein-Smarr
relations do not close with the assigned scaling weights. Within this class, electric
charge and ordinary rotation therefore do not absorb the NUT secondary response, and the
charged rotating family supports
the NUT bi-hair organization: $N$ has a charge-like sector role, while $J_N=MN$
is a rotation-like thermodynamic secondary hair. The latter is defined
off shell in the homogeneous equation of state, not as a new metric parameter or
asymptotic conserved charge.
\end{abstract}

\maketitle

\section{Introduction}

The thermodynamics of NUT-charged spacetimes has remained conceptually subtle since the
original Taub-NUT solution was introduced \cite{AM53-472,JMP4-915,LAM8-160}. The NUT
parameter behaves as a gravitomagnetic charge, is tied to Misner string structures, and
leads to global issues that have no direct analogue in ordinary Kerr-Newman black holes.
These features make the construction of a thermodynamic first law less straightforward
than in more familiar asymptotically flat black hole spacetimes.

Several thermodynamic formulations are now available. One line of
work assigns thermodynamic variables to Misner strings or Misner charges,
treating the string sector as an essential part of the thermodynamic system
\cite{PRD100-064055,JHEP0719119,CQG36-194001,PLB798-134972,PRD100-104016,
JHEP0520084,JHEP1022044,EPJC83-365,EPJC83-589}. Related discussions of line singularities
and dual gravitational charges further emphasize that NUT thermodynamics is sensitive to
variables that are not ordinary electric or rotational charges
\cite{PLB802-135270,PRD106-024022}.
Other approaches modify the thermodynamic energy, use horizon-dependent masses, or
enlarge the thermodynamic state space
\cite{JHEP0321039,PRD101-124011,PRD105-124034,JHEP1022174,IJMPD31-2250021}. In
particular, multi-hair formulations introduce additional thermodynamic variables
adapted to the NUT structure
\cite{PRD100-101501,PRD105-124013,PLB846-138227,2606.03958,PRD108-064034,PRD108-064035}. 
These developments have 
clarified many aspects of NUT thermodynamics, including Lorentzian first laws, charged
and rotating extensions, and NUT anti-de Sitter (AdS) generalizations. At the same time, the coexistence of several consistent first law formulations shows that macroscopic first law
consistency alone does not uniquely determine the thermodynamic state space.

The higher-dimensional charged NUT literature shows the issue clearly.
Komar-type expressions that work in four dimensions can become divergent in higher
even-dimensional NUT geometries, while Abbott-Deser or counterterm methods can provide
finite conserved masses after a suitable background choice. In the multi-hair approach,
the squared-mass relation is interpreted as a hypersurface embedded in a higher
thermodynamic state space, and the physically relevant extension is chosen with the
minimal set of additional hair parameters \cite{PRD108-064034,PRD108-064035}. The connection
between this homogeneous structure and two-horizon sector closure motivates the
diagnostic developed below.

These formulations emphasize complementary aspects of the NUT geometry: the string
sector, the choice of thermodynamic energy, or multiple homogeneous response channels.
Their coexistence motivates the central question of this paper: can two-horizon
thermodynamics discriminate among candidate state spaces once several macroscopic
descriptions are already known to be consistent?
Table~\ref{tab:formulation-comparison} summarizes this separation of questions. The
entries are schematic, but they show how the two-horizon diagnostic complements the
main thermodynamic organizations.

\begin{table*}[t]
\caption{\label{tab:formulation-comparison}
Different NUT thermodynamic organizations address complementary macroscopic questions.
The two-horizon construction tests which candidate response channels close the fixed
sector thermodynamics.}
\centering
\small
\begin{ruledtabular}
\begin{tabular}{p{0.24\textwidth}p{0.33\textwidth}p{0.34\textwidth}}
Organization & Macroscopic Question Answered & Role Relative to the Present Diagnostic \\
Misner string or Misner charge formulations & how string sectors enter a first law &
provides the explicit thermodynamic organization of the string sector \\
modified or horizon mass formulations & which thermodynamic energy gives a consistent
horizon law & provides alternative choices of thermodynamic energy and ensemble \\
multi-hair homogeneous formulations & which homogeneous response channels close mass
formulae and Smarr relations & supplies candidate channels whose sector relevance can be
tested \\
two-horizon sector diagnostic & which candidate response channels are supported by fixed
$\Sigma,\Delta$ sector closure & tests alternatives by simultaneous fixed temperature
first law and Bekenstein-Smarr closure
\end{tabular}
\end{ruledtabular}
\end{table*}

A second ingredient comes from two-horizon black hole thermodynamics and the
black hole/conformal field theory (CFT) correspondence
\cite{LIVING15-11,PLB608-251,JHEP0909088,JHEP0313102,JHEP1112017}.
For black holes with
inner and outer horizons, the horizon entropies and temperatures can be reorganized into
sector variables, often interpreted as left- and right-moving CFT variables. This
thermodynamic method has been useful in Kerr and Kerr-Newman black holes, where the
sector decomposition separates variables that are invisible in one sector from variables
that are needed in the other. Related observations, such as entropy products and
mass-independent horizon combinations, suggest that two-horizon quantities can contain
information not visible in a single-horizon first law
\cite{PRL106-121301,JETP122-113,PLB807-135521,IJMPA38-2350090,IJMPA39-2450044}.

The two-horizon viewpoint avoids changing the macroscopic thermodynamic energy or
assigning a potential to a particular string contribution. The sector entropies and temperatures are fixed
by the inner- and outer horizon thermodynamic quantities. Once these quantities are fixed, the remaining
question is which thermodynamic degrees of freedom allow the sector first laws and Smarr
relations to close. The sector construction is a diagnostic rather than a new
macroscopic convention.
Our contribution is to turn this observation into a state space test. When several
macroscopic thermodynamic formulations are possible, the fixed two-horizon sector
temperatures, together with the sector first laws and Smarr relations, constrain the
candidate thermodynamic degrees of freedom. A viable state space must simultaneously
reproduce the first law differential, the sector temperature fixed by the horizon
quantities, and the corresponding Bekenstein-Smarr relation. This separates response channels
even when several ordinary first laws are separately consistent.

The charged rotating case tests this question beyond the uncharged prototype. Once the
ordinary electric charge $Q$ and the ordinary angular momentum
$J=Ma$, where $a$ is the rotation parameter, are present, the NUT secondary hair might
in principle become unnecessary. The
$\Delta$ sector could have closed with $(M,Q,N,J)$ alone, or the effects of $J_N$ could
have been absorbed into the usual Kerr-Newman variables. Either outcome would have
weakened the bi-hair interpretation. The Kerr-Newman-NUT geometry is a stress
test of the diagnostic, not merely a longer version of the uncharged calculation.

In the uncharged Taub-NUT case this diagnostic gives a clean result
\cite{2607.23644}. The sum sector is described by the mass and the NUT charge
$N$, whereas the difference sector closure test favors including the rotation-like
thermodynamic secondary hair $J_N=MN$. This reveals a bi-hair thermodynamic
organization through a charge-like facet $N$ and a rotation-like facet $J_N$. The
secondary hair is defined off shell in the homogeneous equation of state, analogous in
spirit to treating $J=Ma$ as an independent thermodynamic variable for rotating black
holes; it is not an additional metric parameter or an independent asymptotic Noether
charge.

For the homogeneous state spaces defined and tested below, we find that the charged
extension preserves the same organizing principle found in the uncharged case. The sum
sector of Kerr-Newman-NUT thermodynamics is controlled by
$M$, $N$, and $Q$, while the difference sector additionally requires the
rotation-like variables $J$ and $J_N$. The
Christodoulou-Ruffini-type mass formula
\cite{PRL25-1596,PRD4-3552}, entropy product, sector first laws, and sector Smarr
relations all take homogeneous forms in this enlarged state space. Reduced state spaces
tested below lose the sector temperature fixed by the horizons, a first law differential, or a
Bekenstein-Smarr term. In the appropriate limits, the construction reduces to the Kerr-Newman sector
thermodynamics when $N\to0$, and to the Taub-NUT bi-hair result when $Q,a\to0$.

Compared with the uncharged prototype, the charged problem forces several distinctions.
The electric charge is a
genuine conserved charge and it enters both sectors, so it must be separated from the
sector role of the NUT parameter; the statement that $N$ is charge-like is a statement
about sector thermodynamics, not about a Maxwell-type gauge charge. Ordinary
rotation and NUT secondary hair appear together in the difference sector, making it
possible to test whether they are independent response channels. The charged
Christodoulou-Ruffini-type mass formula contains $Q$, $J$, $N$, and $J_N$ in one
homogeneous equation of state, so the proposed state space can be checked against a
single algebraic relation.

We use geometric units in which Newton's constant $G$, the speed of light $c$, the
reduced Planck constant $\hbar$, and the Boltzmann constant $k_{\rm B}$ are set to one.
Our paper is organized as follows. Section~\ref{sec:sector-diagnostic} formulates the
two-horizon sector diagnostic. Section~\ref{sec:tn-prototype} reviews the uncharged
Taub-NUT case as the reference sector test underlying the present extension.
Section~\ref{sec:benchmarks} reviews Kerr and Kerr-Newman as benchmarks.
Section~\ref{sec:knnut-thermo} presents the horizon thermodynamics of Kerr-Newman-NUT
spacetime in the enlarged state space. Section~\ref{sec:mass-product} derives the
corresponding mass formula and entropy product. Section~\ref{sec:cft-sectors}
constructs the sum and difference sectors and derives their first laws and Smarr
relations. Section~\ref{sec:sector-roles} summarizes the sector roles and limiting
checks. Section~\ref{sec:hidden-cft} gives a rotating parent hidden CFT normalization
check.
Section~\ref{sec:discussion} discusses the physical meaning of the result and possible
extensions.

\section{Two-horizon sector diagnostic}
\label{sec:sector-diagnostic}

We first summarize the diagnostic in a form that can be applied to Kerr, Kerr-Newman,
Taub-NUT, and Kerr-Newman-NUT without committing to a particular microscopic CFT
realization \cite{PLB608-251,JHEP0909088,JHEP0313102}. Consider a stationary spacetime
with outer and inner horizons, entropies $S_\pm$, and signed temperatures $T_\pm$. We
define
\begin{equation}\label{Spm}
S_{\Sigma}=\frac{S_+ +S_-}{2},\qquad
S_{\Delta}=\frac{S_+ -S_-}{2},
\end{equation}
and
\begin{equation}\label{Tpm}
\frac{1}{T_{\Sigma,\Delta}}=\frac{1}{T_+}\pm\frac{1}{T_-}.
\end{equation}
Throughout this paper the inner horizon temperature $T_-$ is the signed thermodynamic
temperature. The labels $\Sigma$ and $\Delta$ are used instead
of left and right because the assignment of microscopic $L/R$ names can be
convention dependent, whereas the sum and difference sectors are unambiguous.

The diagnostic combines three requirements. Horizon data fix the sector entropy and
temperature. The allowed extensive variables must then close both the first law and its
Bekenstein-Smarr relation. The coefficient fixed by the horizon temperature is the main
discriminator. For a homogeneous equation of state, the Bekenstein-Smarr relation is the
corresponding Euler consistency condition, not an independent empirical test.
The order of these steps is essential. The horizons determine the sector temperature
before any conjugate potentials are introduced; the first law and Smarr relation then
test the proposed state space. If the potentials were allowed to define the temperature,
the construction would lose its ability to distinguish different state spaces.

This is more restrictive than rewriting $S_+$ and $S_-$ in different parameters on the
physical solution surface. Once the two horizons have fixed the sector temperature, its
coefficient cannot be changed by renaming a conjugate potential. Removing a candidate
variable therefore tests thermodynamic content rather than notation.
We use ``state space'' in the usual thermodynamic sense. For example, $J=Ma$ is treated
as an extensive variable even when a rotating metric is parametrized by $M$ and $a$.
Likewise, $J_N=MN$ is built from metric parameters on the physical family, while partial
derivatives are taken in an enlarged homogeneous state space.

An off-shell extension is therefore needed to define the independent response
coefficients, but it cannot be chosen arbitrarily. We use the following restrictions to
keep that extension tied to the physical horizon thermodynamics.
An \emph{admissible homogeneous state space} is a minimal extension whose variables have
definite scaling weights, have a physical origin in the horizon differential, the
homogeneous entropy polynomial, or a surface-charge construction, and reduce correctly
in the Kerr-Newman and Taub-NUT limits. They must not vanish identically on the physical
submanifold. These conditions are imposed before testing sector closure.

With this definition in place, the horizon differential, the mass formula, and available
surface-charge constructions determine the candidate list.
The candidate variables are the weight one charges $Q$ and $N$ and the weight two
variables $J$ and $J_N$. The ordinary angular momentum $J$ is tied to the axial Killing
generator. The NUT secondary hair $J_N$ has a background-subtracted Komar-type surface
integral realization in the multihair construction, where it was proposed as a
subleading dual charge \cite{PRD108-064035}; its complete asymptotic interpretation
remains open. The horizon differential and the Christodoulou-Ruffini-type mass formula
also supply distinct responses for these variables. By contrast, $N^2$ is generated by
$N$, while $MQ$ and $QN$ have no independent conjugate coefficients in the
thermodynamics considered here. The same rule retains $J$ in Kerr and Kerr-Newman.
This candidate list is the input to the diagnostic, not its conclusion. After
$S_{\Sigma,\Delta}$ and $T_{\Sigma,\Delta}$ have been fixed, we remove candidate
channels and check whether the first law and Smarr relation still close. Thus admitting
$J_N$ because it has the same scaling weight as $J$ does not assume the bi-hair result.
Its role is decided by the reduced state spaces examined below.

The same minimality requirement controls how the physical family is continued off shell.
The on-shell horizon entropy does not determine a unique continuation: terms that vanish
when $J_N=MN$ is imposed can still change off-shell derivatives. We do not attempt to
classify all such continuations. Instead, we restrict the comparison to the
lowest-degree polynomial homogeneous extension built from the physically identified
candidate channels, with the correct Kerr-Newman and Taub-NUT limits. Any residual
coefficient in this class is displayed explicitly and tested against the temperature
fixed by the horizons. Arbitrary higher-order functions that vanish on the physical
submanifold would introduce unconstrained response data and lie outside the stated
diagnostic class. This restriction follows the embedding viewpoint used for multihair
mass formulae in higher dimensional NUT systems
\cite{PRD108-064034,PRD108-064035}.
Homogeneity supplies the final restriction. Charges such as $Q$ and $N$ have weight one,
whereas $J$ and $J_N$ have weight two. The sector Smarr relations are the corresponding
Euler relations, so both the variables and their coefficients must have the correct
weights.

The diagnostic can now be applied in a fixed order: the horizon and mass data supply
candidate variables, the two horizons fix the sector temperature, and reduced state
spaces are tested for first law and Smarr closure.
For a sector with energy $E_\alpha=M/2$ and extensive variables $X^A_\alpha$, the required structure is
\begin{equation}
dE_\alpha=T_\alpha dS_\alpha+\sum_A \mu^\alpha_A dX^A_\alpha ,
\end{equation}
where $\alpha=\Sigma,\Delta$ labels the sector, $A$ labels its extensive variables, and
$\mu^\alpha_A$ denotes the potential conjugate to $X^A_\alpha$. A compatible Smarr
relation follows from homogeneous scaling. Which variables
belong in the list $X^A_\alpha$ is the question being tested. If a proposed set of
variables cannot reproduce the fixed $T_\alpha$, or if its scaling assignment fails the
corresponding Smarr relation, it is not the sector state space identified by this
diagnostic within the restricted homogeneous class.

The benchmark cases motivate testing whether the $\Sigma$ sector remains charge-like and
whether the $\Delta$ sector remains sensitive to rotation-like responses when $N$ and
$J_N$ are present. We do not assume these assignments in advance. They are inferred from
the closure of the first laws and Smarr relations. This matters for NUT thermodynamics,
where the same geometric parameter can have both charge-like and rotation-like
thermodynamic facets.

\section{Uncharged Taub-NUT as a reference sector test}
\label{sec:tn-prototype}

Before turning to the charged system, we recall previous analysis \cite{2607.23644} for the uncharged Taub-NUT case, which is  the simplest version of the
sector diagnostic. Here we review only the ingredients needed to make the charged
generalization self contained. The uncharged case is the cleanest
setting because the only macroscopic parameters are the mass scale and the NUT parameter,
while the two-horizon sector construction still distinguishes two different
thermodynamic responses. It is the reference case for the longer charged analysis:
all new structures in the Kerr-Newman-NUT case should reduce either to the usual
Kerr-Newman sector thermodynamics or to this Taub-NUT sector pattern in the appropriate
limits.

The outer and inner horizon radii are
\begin{equation}
r_\pm=M\pm\sqrt{M^2+N^2},
\end{equation}
the horizon entropies and signed temperatures are
\begin{equation}
S_\pm=\pi(r_\pm^2+N^2),\qquad
T_\pm=\frac{r_\pm-M}{2\pi(r_\pm^2+N^2)}.
\end{equation}
The multi-hair horizon thermodynamics is obtained by introducing the charge-like NUT
variable $N$ and the secondary hair
\begin{equation}
J_N=MN.
\end{equation}
The horizon first law and Smarr relation take the standard homogeneous form
\begin{equation}
dM=T_\pm dS_\pm+\omega_\pm dJ_N+\phi_\pm dN,
\end{equation}
\begin{equation}
M=2(T_\pm S_\pm+\omega_\pm J_N)+\phi_\pm N,
\end{equation}
where $\omega_\pm$ and $\phi_\pm$ are conjugate to $J_N$ and $N$, respectively:
\begin{equation}
\omega_\pm=\frac{N}{r_\pm^2+N^2},\qquad
\phi_\pm=-\frac{2Nr_\pm}{r_\pm^2+N^2}.
\end{equation}
These equations display two NUT thermodynamic responses:
one is conjugate to $N$, and the other is conjugate to the rotation-like combination
$J_N$. The point of the sector diagnostic is to test whether this split is merely one
possible macroscopic convention or whether it is favored by the two-horizon sector
structure within the restricted homogeneous class.

For a generic horizon entropy $S$, the same candidate homogeneous state space admits a
Christodoulou-Ruffini-type mass formula,
\begin{equation}
M^2=\frac{\pi}{S}\left(\frac{S}{2\pi}-N^2\right)^2
    +\frac{\pi J_N^2}{S}.
\end{equation}
It gives the thermodynamic potentials by partial differentiation in the enlarged
homogeneous state space. The physical Taub-NUT family is recovered after imposing
$J_N=MN$, but the derivatives defining $\omega_\pm$ are taken before this restriction is
made, just as derivatives with respect to $J=Ma$ are used for rotating black holes whose
metrics are written in terms of $M$ and $a$.

The entropy product also has a simple homogeneous form,
\begin{equation}
S_+S_-=4\pi^2(N^4+J_N^2).
\end{equation}
We do not regard this product as a proof of microscopic universality, since on the
physical submanifold $J_N=MN$ it still carries the mass dependence through $J_N$.
Instead, it is a structural clue: the same pair $(N,J_N)$ that appears in the horizon
first law also organizes the two-horizon entropy quantities.

The sector quantities fixed by the two horizons are
\begin{equation}
S_{\Sigma}=2\pi(M^2+N^2),\qquad
T_{\Sigma}=\frac{1}{8\pi M},
\end{equation}
and
\begin{equation}
S_{\Delta}=2\pi M\sqrt{M^2+N^2},\qquad
T_{\Delta}=\frac{1}{8\pi\sqrt{M^2+N^2}}.
\end{equation}

The sum sector closes in the charge-like state space generated by $M$ and $N$. With
$E_{\Sigma}=M/2$ and $N_{\Sigma}=N/2$, one obtains
\begin{equation}
dE_{\Sigma}=T_{\Sigma}dS_{\Sigma}+\phi_{\Sigma}dN_{\Sigma},
\qquad
\phi_{\Sigma}=-\frac{N}{M},
\end{equation}
together with the corresponding Smarr relation. The sector construction does not
replace the NUT charge by a rotation-like variable: the charge-like role of $N$ is
already visible in the sum sector.

The difference sector is more restrictive. If one describes it only by $M$ and $N$, the
entropy differential is
\begin{equation}
\frac{dS_{\Delta}}{2\pi}
=\frac{(2M^2+N^2)dM+MN\,dN}{\sqrt{M^2+N^2}},
\end{equation}
and the coefficient of $dM$ is incompatible with the fixed temperature
$T_{\Delta}=1/(8\pi\sqrt{M^2+N^2})$. The failure cannot be repaired by changing only the
potential conjugate to $N$, because the temperature is already fixed by the horizon
harmonic mean. If instead one tries to describe the sector only by $M$ and
$J_N=MN$, one may write
\begin{equation}
S_{\Delta}=2\pi\sqrt{M^4+J_N^2},
\end{equation}
but the resulting derivative with respect to $M$ again fails to reproduce the same
$T_{\Delta}$. Within this restricted homogeneous extension, these two failures indicate
that the difference sector is sensitive to two distinct NUT responses: the charge-like
response associated with $N$ and the
rotation-like response associated with $J_N$.

The minimal homogeneous sector extension then uses the pair $(N,J_N)$. We write
\begin{equation}
S_{\Delta}
=2\pi\sqrt{M^4+(1+w)M^2N^2-wJ_N^2},
\end{equation}
where $w$ is a dimensionless coefficient to be fixed by the sector first law. This
lowest-degree polynomial extension uses the mass scale, the NUT charge, and the
rotation-like combination already present in the horizon thermodynamics and the mass
formula. Its $w$ dependence is proportional to $M^2N^2-J_N^2$ and therefore vanishes on
the physical submanifold. We keep this residual ambiguity explicit instead of choosing
an off-shell continuation in advance.

Taking the differential and then restricting to the physical submanifold gives
\begin{align}
\frac{dS_{\Delta}}{2\pi}
=&
\frac{[2M^2+(1+w)N^2]dM}{\sqrt{M^2+N^2}}
\nonumber\\
&+\frac{(1+w)MN\,dN-wN\,dJ_N}{\sqrt{M^2+N^2}}.
\end{align}
The potentials multiplying $dN$ and $dJ_N$ cannot alter the coefficient of $dM$. The
sector temperature fixed by the horizons therefore gives $w=1$. The resulting first law is
\begin{equation}
dE_{\Delta}
=T_{\Delta}dS_{\Delta}
+\omega_{\Delta}dJ_N+\phi_{\Delta}dN_{\Delta},
\end{equation}
where $E_\Delta=M/2$, $N_\Delta=N/2$, and
\begin{equation}
\omega_{\Delta}=\frac{N}{4(M^2+N^2)},\qquad
\phi_{\Delta}=-\frac{MN}{M^2+N^2},
\end{equation}
and it obeys the Bekenstein-Smarr relation
\begin{equation}
E_{\Delta}
=2T_{\Delta}S_{\Delta}+2\omega_{\Delta}J_N+\phi_{\Delta}N_{\Delta}.
\end{equation}

The uncharged calculation thus assigns two thermodynamic roles to the NUT parameter:
$N$ is charge-like, while $J_N=MN$ is a rotation-like secondary hair. The latter is
neither a new metric parameter nor an asymptotic Noether charge. Its inclusion closes the
difference sector first law and Smarr relation in the restricted homogeneous state space.
The charged analysis below tests whether this separation survives in the presence of
electric charge and ordinary angular momentum.

\section{Benchmarks: Kerr and Kerr-Newman}
\label{sec:benchmarks}

We use two benchmark cases before introducing the NUT charge. The
Kerr solution shows the basic pattern of the sector diagnostic: the sum sector is blind
to rotation, whereas the difference sector carries the angular momentum. The
Kerr-Newman solution then shows how an ordinary electric charge enters the same
two-horizon sector structure. These examples set the reference point for the
Kerr-Newman-NUT analysis below, where both the ordinary rotation $J$ and the NUT
secondary hair $J_N$ are present. They also serve as checks on the normalization of the
sector energy and charge variables. Throughout this paper the sector energy is
$E_\alpha=M/2$, and ordinary charge variables are split as $Q_\alpha=Q/2$ when they
appear in a sector first law.

The benchmark construction is closely related to the thermodynamic reformulation of
Kerr-Newman black holes in Ref. \cite{PLB608-251}. There one forms two effective
systems from the inner and outer horizons by taking the sum and difference of the
reduced horizon areas $A_\pm=S_\pm/\pi$, namely $A_R=A_++A_-$ and
$A_L=A_+-A_-$. The relevant feature is that the
system associated with the sum carries only $(M,Q)$, while the one associated with the
difference carries $(M,Q,J)$. Their temperatures are related to the inner and outer
horizon temperatures by harmonic combinations, which motivates the sector language used here. Our notation is as in \eqref{Spm} and \eqref{Tpm}, 
together with sector energy
$M/2$, so that the sector first laws take the same form as the later NUT sector
relations, rendering  the sector Smarr formulae parallel to one another. 
The invariant
content is the sum/difference decomposition itself: one sector is insensitive to
rotation, while the other detects rotation-like thermodynamic variables. We test this
pattern in NUT-charged geometries.

\subsection{Kerr}

The Kerr metric, with rotation parameter $a$, angular momentum $J=Ma$, and horizon radii
\begin{equation}
r_\pm=M\pm\sqrt{M^2-a^2},
\end{equation}
has horizon entropies and signed temperatures  
\begin{equation}
S_\pm=2\pi Mr_\pm,\qquad
T_\pm=\frac{r_\pm-M}{4\pi Mr_\pm}.
\end{equation}
The sector quantities are
\begin{equation}
S_{\Sigma}=2\pi M^2,\qquad
T_{\Sigma}=\frac{1}{8\pi M},
\end{equation}
and
\begin{equation}
S_{\Delta}=2\pi\sqrt{M^4-J^2},\qquad
T_{\Delta}=\frac{\sqrt{M^2-a^2}}{8\pi M^2}.
\end{equation}
For Kerr, the sum sector closes with $E_{\Sigma}=M/2$ alone,
\begin{equation}
dE_{\Sigma}=T_{\Sigma}dS_{\Sigma},\qquad
E_{\Sigma}=2T_{\Sigma}S_{\Sigma},
\end{equation}
whereas the difference sector needs $J$,
\begin{equation}
dE_{\Delta}=T_{\Delta}dS_{\Delta}+\Omega_{\Delta}dJ,\qquad
E_{\Delta}=2T_{\Delta}S_{\Delta}+2\Omega_{\Delta}J,
\end{equation}
where $\Omega_\Delta$ is the angular potential,
\begin{equation}
\Omega_{\Delta}=\frac{a}{4M^2}.
\end{equation}
The entropy difference is the rotation benchmark. It detects $J$, as expected from the
Kerr equation of state, and motivates testing whether $J_N=MN$ plays the same role in
the NUT difference sector.
The Kerr limit also illustrates the thermodynamic meaning of extremality in the sector
language. When $a\to M$, the difference sector entropy and temperature both vanish,
whereas the sum sector remains finite. The sector sensitive to rotation is the one
that detects the approach to the extremal rotating state. This behavior persists in the
Kerr-Newman benchmark, with the extremal condition shifted to $M^2=Q^2+a^2$, and it is
one reason to look for rotation-like NUT responses in the difference sector.

\subsection{Kerr-Newman}

For Kerr-Newman, the horizon radii are
\begin{equation}
r_\pm=M\pm\sqrt{M^2-Q^2-a^2},
\end{equation}
\begin{equation}
S_\pm=\pi(2Mr_\pm-Q^2),\qquad
T_\pm=\frac{r_\pm-M}{2\pi(r_\pm^2+a^2)}.
\end{equation}
Equivalently,  the sum and
difference systems are  \cite{PLB608-251}
\begin{align}
S_{\Sigma}&=\pi \left(2M^2-Q^2\right),\\
S_{\Delta}& =2\pi M\sqrt{M^2-Q^2-a^2}.
\end{align}
where the factor of one half is accompanied by the sector energy
$M/2$, so the thermodynamic content is the same.

In the sum sector 
\begin{equation}
T_{\Sigma}=\frac{1}{8\pi M},\qquad
\Phi_{\Sigma}=\frac{Q}{2M}
\end{equation}
where $\Phi_\Sigma$ is the electric potential of the sum sector. We obtain
\begin{equation}
\frac{dM}{2}=T_{\Sigma}dS_{\Sigma}+\Phi_{\Sigma}\frac{dQ}{2}.
\end{equation}
The sum sector is sensitive to the charge-like variable $Q$, but remains blind to
the angular momentum. Unlike rotation in Kerr, a gauge charge contributes
to the entropy sum and to the sector whose temperature is
$1/(8\pi M)$. The associated Smarr relation is
\begin{equation}
\frac{M}{2}=2T_{\Sigma}S_{\Sigma}+\Phi_{\Sigma}\frac{Q}{2}.
\end{equation}
In the language of Ref. \cite{PLB608-251}, the sum system behaves as a nonrotating
charged black hole: it carries $M$ and $Q$, but it
does not carry the macroscopic angular momentum.

The difference sector is
\begin{equation}
S_{\Delta}=2\pi\sqrt{M^4-M^2Q^2-J^2},
\end{equation}
\begin{equation}
T_{\Delta}=\frac{\sqrt{M^2-Q^2-a^2}}{4\pi(2M^2-Q^2)},
\end{equation}
with angular and electric potentials
\begin{equation}
\Omega_{\Delta}=\frac{a}{2(2M^2-Q^2)},\qquad
\Phi_{\Delta}=\frac{MQ}{2M^2-Q^2}.
\end{equation}
The first law takes the sector form
\begin{equation}
\frac{dM}{2}
=T_{\Delta}dS_{\Delta}
+\Omega_{\Delta}dJ+\Phi_{\Delta}\frac{dQ}{2} 
\end{equation}
along with the Smarr-type relation
\begin{equation}
\frac{M}{2}
=2T_{\Delta}S_{\Delta}
+2\Omega_{\Delta}J+\Phi_{\Delta}\frac{Q}{2}.
\end{equation}
The difference system  retains the electric charge and also
carries the angular momentum  \cite{PLB608-251}. This asymmetry motivates the NUT analysis: a response with
angular momentum scaling may be absent from the sum sector yet required by the
difference sector.

The harmonic relations take the expected form. With the signed inner horizon
temperature,
\begin{equation}
\frac{1}{T_{\Sigma}}=\frac{1}{T_+}+\frac{1}{T_-},\qquad
\frac{1}{T_{\Delta}}=\frac{1}{T_+}-\frac{1}{T_-}.
\end{equation}
The sector temperatures are not chosen to make the first laws work; they are fixed
by the two horizon temperatures before the state space is tested.

The Kerr-Newman metric already gives the separation needed for comparison: the electric charge enters
both sectors, while the rotation-like variable $J$ is needed by the difference
sector. The charged NUT system studied below should reduce to this benchmark when
$N\to0$, while for $Q,a\to0$ it should reduce to the Taub-NUT reference result of
Sec.~\ref{sec:tn-prototype}.
Adding electric charge changes more than a single potential in the Taub-NUT construction.
In Kerr-Newman, $Q$ appears in the  entropy, the  
temperature, and the electromagnetic potential in both sectors. Once a NUT
charge is also present, the sector diagnostic must decide how the ordinary charge
sector, the ordinary rotation sector, and the two NUT responses coexist in a single
homogeneous state space.

\section{Kerr-Newman-NUT thermodynamics}
\label{sec:knnut-thermo}

We now turn to the charged and rotating NUT geometry. To fix conventions, we use the
standard Lorentzian Kerr-Newman-NUT metric, with signs chosen consistently with the
previous conventions \cite{JHEP0520084,PRD105-124013} for 
NUT thermodynamics:  
\begin{align}
ds^2={}&-\frac{\Delta}{\rho^2}\left(dt-P\,d\varphi\right)^2
 +\frac{\rho^2}{\Delta}dr^2+\rho^2d\theta^2 \nonumber\\
&+\frac{\sin^2\theta}{\rho^2}
\left[a\,dt-(r^2+a^2+N^2)d\varphi\right]^2,
\end{align}
where $(t,r,\theta,\varphi)$ are Boyer-Lindquist-type coordinates, while $M$, $a$, $Q$,
and $N$ denote the mass, rotation parameter, electric charge, and NUT charge,
respectively. The metric functions are
\begin{equation}
\rho^2=r^2+(N+a\cos\theta)^2,\qquad
P=a\sin^2\theta-2N\cos\theta,
\end{equation}
and
\begin{equation}
\Delta=r^2-2Mr+a^2+Q^2-N^2.
\end{equation}
Different sign conventions for the NUT parameter appear in the literature; all
thermodynamic quantities below refer to the convention displayed here.
The electromagnetic gauge potential one-form $A$ may be chosen as
\begin{equation}
A=-\frac{Qr}{\rho^2}\left(dt-P\,d\varphi\right),
\end{equation}
up to the usual gauge shifts. The electric potentials below are evaluated in this gauge
with respect to the horizon generators. Here $Q$ is normalized as the ordinary Maxwell
electric charge measured by its asymptotic flux. On either horizon,
$1-\Omega_\pm P=\rho_\pm^2/(r_\pm^2+a^2+N^2)$, where $\Omega_\pm$ are
the horizon angular velocities.
The horizon
generators are $\chi_\pm=\partial_t+\Omega_\pm\partial_\varphi$, and
so we find that
$-\chi_\pm\mathbin{\cdot}A=Qr_\pm/(r_\pm^2+a^2+N^2)$ is independent of $\theta$.
Possible patch choices associated with the
string do not affect the sector thermodynamic identities considered here. With this
convention, the radial function used below is
$f(r)=\Delta$,
and the two Killing horizons are located at
\begin{equation}
r_\pm=M\pm\sqrt{M^2+N^2-Q^2-a^2}.
\end{equation}
where $M^2+N^2\ge Q^2+a^2$ is required to ensure 
the existence of two real horizons.

The horizon
thermodynamic quantities are
\begin{equation}
S_\pm=\pi(r_\pm^2+a^2+N^2)
     =\pi(2Mr_\pm+2N^2-Q^2),
\end{equation}
\begin{equation}
T_\pm=\frac{r_\pm-M}{2\pi(r_\pm^2+a^2+N^2)}.
\end{equation}
$T_-$ is the signed temperature used in the sector construction. With
this convention the algebraic relations between inner and outer horizons remain
parallel, and the sum/difference sector temperatures are obtained directly from the
harmonic combinations of $T_+$ and $T_-$.

We adopt the multi-hair homogeneous representation as a candidate off-shell extension.
Its extensive variables include the ordinary angular momentum and the NUT secondary hair,
\begin{equation}
J=Ma,\qquad J_N=MN.
\end{equation}
Here $J_N$ is introduced in the same thermodynamic sense as in the uncharged prototype: as before, 
it is not an additional metric parameter but a homogeneous thermodynamic secondary hair built from the
mass scale and the NUT parameter. A background-subtracted Komar-type surface integral
provides an independent construction of $J_N$ as a candidate subleading dual charge
\cite{PRD108-064035}, although its complete asymptotic charge interpretation remains an
open question. The metric family is still parametrized by
$(M,a,Q,N)$; the enlarged state space is a candidate homogeneous mass representation that allows
partial derivatives to test distinct charge-like and rotation-like NUT responses. With
this choice, the horizon equation can be solved for the mass as
\begin{equation}
M=\frac{r_h}{2}+\frac{a^2+Q^2-N^2}{2r_h},
\end{equation}
where $r_h$ denotes either $r_+$ or $r_-$, and $S_h$ denotes the corresponding horizon entropy. The differential of this relation, together with the identities
\begin{equation}
S_\pm=\pi(r_\pm^2+a^2+N^2),\qquad
J=Ma,\qquad J_N=MN,
\end{equation}
is the direct way to obtain the horizon first law. The calculation is elementary but
instructive. One first uses
\begin{equation}
dJ=a\,dM+M\,da,\qquad dJ_N=N\,dM+M\,dN,
\end{equation}
to replace variations of $a$ and of the product $MN$ by thermodynamic variations.
Solving the differential of the horizon equations $\Delta(r_\pm)=0$ for $dM$ then gives
\begin{align}
dM={}&\frac{r_\pm-M}{2\pi D_\pm}dS_\pm
+\frac{a}{D_\pm}dJ+\frac{Qr_\pm}{D_\pm}dQ \nonumber\\
&+\frac{N}{D_\pm}dJ_N-\frac{2Nr_\pm}{D_\pm}dN,
\end{align}
where
\begin{equation}
D_\pm=r_\pm^2+a^2+N^2 .
\end{equation}
Variations of $a$ are rewritten in terms of $dJ$, while variations of the product $MN$
are rewritten in terms of $dJ_N$ and $dN$. In this candidate representation the horizon
differential therefore supplies two NUT conjugate coefficients. This rewriting
does not by itself establish that $J_N$ is an independent physical channel; its
diagnostic relevance is tested below against reduced admissible embeddings.
The outer and inner horizons then obey a common first law structure,

\begin{equation}
dM=T_\pm dS_\pm+\Omega_\pm dJ+\Phi_\pm dQ+\omega_\pm dJ_N+\phi_\pm dN,
\end{equation}
and a compatible Smarr relation,
\begin{equation}
M=2(T_\pm S_\pm+\Omega_\pm J+\omega_\pm J_N)+\Phi_\pm Q+\phi_\pm N,
\end{equation}
where $\Omega_\pm$, $\Phi_\pm$, $\omega_\pm$, and $\phi_\pm$ are conjugate to
$J$, $Q$, $J_N$, and $N$, respectively:
\begin{equation}
\Omega_\pm=\frac{a}{r_\pm^2+a^2+N^2},\quad
\Phi_\pm=\frac{Qr_\pm}{r_\pm^2+a^2+N^2},
\end{equation}
\begin{equation}
\omega_\pm=\frac{N}{r_\pm^2+a^2+N^2},\quad
\phi_\pm=-\frac{2Nr_\pm}{r_\pm^2+a^2+N^2}.
\end{equation}
Equivalently, using the horizon identity
\begin{equation}
r_\pm^2+a^2+N^2=2Mr_\pm+2N^2-Q^2,
\end{equation}
all potentials can be written directly in terms of the horizon radii and the
thermodynamic parameters. This form is convenient when taking sums and differences of
the inner- and outer horizon quantities.
The pair $(\omega_\pm,\phi_\pm)$ is the charged and rotating analogue of the two candidate
NUT responses seen in the Taub-NUT prototype. In the enlarged representation, the first
coefficient is conjugate to $J_N$ and the second to $N$. The horizon thermodynamics
therefore supplies candidate response channels; the subsequent two-horizon sector
construction tests which of them are supported by the $\Sigma$ and $\Delta$ first laws
and Smarr relations.

The variable $J_N$ is a thermodynamic secondary hair, not a metric parameter or an
independent conserved charge at infinity. The physical Kerr-Newman-NUT family is the
submanifold $J_N=MN$. A derivative with respect to $J_N$ probes an off-shell,
weight two NUT response in the homogeneous equation of state; it is not a path through
new metrics. The physical relation is imposed after differentiation. Accordingly,
$\omega_\pm$ is a response coefficient rather than the chemical potential of a new
asymptotic Noether charge. The present diagnostic does not vary the Misner string
independently and uses only sector quantities fixed by the horizons.

Several immediate limits are worth recording. If $N\to0$, then $J_N$, $\omega_\pm$, and
$\phi_\pm$ disappear and the standard Kerr-Newman horizon first law is recovered. If
$Q\to0$ while $a$ is kept, the formulae describe the rotating Kerr-NUT sector of the
same state space. If both $Q$ and $a$ are sent to zero, the horizon first law reduces to
the uncharged Taub-NUT multi-hair form reviewed in Sec.~\ref{sec:tn-prototype}. These
limits confirm that the candidate representation has the expected Kerr-Newman and
Taub-NUT reductions.

\section{Mass formula and entropy product}
\label{sec:mass-product}

The same candidate homogeneous state space is encoded in a Christodoulou-Ruffini-type equation of state. For a
generic horizon entropy $S$, the mass satisfies
\begin{equation}
M^2=\frac{\pi}{S}\left(\frac{S/\pi+Q^2-2N^2}{2}\right)^2
    +\frac{\pi(J^2+J_N^2)}{S}.
\end{equation}
Here $(S,J,J_N,Q,N)$ are thermodynamic variables with definite scaling weights.
The entropy has weight two; $M$, $Q$, and $N$ have
weight one; and $J$ and $J_N$ have weight two. Euler scaling of this equation of state
is the origin of the Smarr relations above and below. Equivalently, the horizon
entropies are the two roots of
\begin{align}
0={}&S^2-4\pi\left(M^2+N^2-\frac{Q^2}{2}\right)S \nonumber\\
&+4\pi^2\left[J^2+J_N^2+\left(N^2-\frac{Q^2}{2}\right)^2\right].
\end{align}
The entropy sum and product are
\begin{equation}
S_+ +S_-=4\pi\left(M^2+N^2-\frac{Q^2}{2}\right),
\end{equation}
\begin{equation}
S_+S_-=4\pi^2\left[J^2+J_N^2+\left(N^2-\frac{Q^2}{2}\right)^2\right].
\end{equation}
The product is explicitly independent of $M$ only when written in the extended
homogeneous variables. On the physical Kerr-Newman-NUT submanifold one must still impose
$J=Ma$ and $J_N=MN$, so the product is not being used as a separate proof of
mass independence at fixed metric parameters. Its role here is to display the same
homogeneous variables that organize the first laws.
One can also see directly how the horizon first law follows from the mass formula. If
$M=M(S,J,J_N,Q,N)$, its differential is
\begin{equation}
dM=T\,dS+\Omega\,dJ+\omega\,dJ_N+\Phi\,dQ+\phi\,dN,
\end{equation}
with
\begin{equation}
T=\left(\frac{\partial M}{\partial S}\right)_{J,J_N,Q,N},\quad
\Omega=\left(\frac{\partial M}{\partial J}\right)_{S,J_N,Q,N},
\end{equation}
\begin{align}
\omega&=\left(\frac{\partial M}{\partial J_N}\right)_{S,J,Q,N},\\
\Phi&=\left(\frac{\partial M}{\partial Q}\right)_{S,J,J_N,N},\\
\phi&=\left(\frac{\partial M}{\partial N}\right)_{S,J,J_N,Q}.
\end{align}
After substituting $S=S_\pm$, $J=Ma$, and $J_N=MN$, these derivatives reproduce the
horizon potentials displayed in Sec.~\ref{sec:knnut-thermo}. The horizon first law and the
Christodoulou-Ruffini-type equation of state are not independent assumptions; they are
two equivalent ways of encoding the same enlarged thermodynamic state space.

The product is not used here as an independent proof of microscopic universality.
Instead, it shows that the candidate homogeneous variables appearing in the horizon first
laws also organize the two-horizon quantities. The appearance of $J_N$
alongside the ordinary angular momentum $J$ motivates testing whether it supplies a
distinct rotation-like response.
The entropy polynomial anticipates a possible sector decomposition. The coefficient of
$S$ determines the entropy sum and hence the $\Sigma$ sector; it contains the
charge-like variables $N$ and $Q$ but not the angular momenta. The constant term
contains $J^2+J_N^2$ and tracks both candidate rotation-like variables. It therefore
suggests a charge-like sum sector and a difference sector sensitive to $J$ and $J_N$,
but it does not determine the sector state space. The test comes from the sector first
laws and Smarr relations, to which we now turn.

\section{Thermodynamic sum and difference sectors of Kerr-Newman-NUT}
\label{sec:cft-sectors}

We now apply the sector definitions of Sec.~\ref{sec:sector-diagnostic}. The entropy sum gives
\begin{equation}
S_{\Sigma}=2\pi\left(M^2+N^2-\frac{Q^2}{2}\right),
\end{equation}
and the harmonic mean temperature is
\begin{equation}
T_{\Sigma}=\frac{1}{8\pi M},
\end{equation}
which is independent of $a$ and of the NUT secondary hair. Differentiating
$S_{\Sigma}$ gives the sector first law
\begin{equation}
\frac{dM}{2}=T_{\Sigma}dS_{\Sigma}
-\frac{N}{2M}dN+\frac{Q}{4M}dQ.
\end{equation}
Equivalently, if $N_{\Sigma}=N/2$ and $Q_{\Sigma}=Q/2$, then
\begin{equation}
\phi_{\Sigma}=-\frac{N}{M},\qquad
\Phi_{\Sigma}=\frac{Q}{2M},
\end{equation}
yielding
\begin{equation}
\frac{dM}{2}=T_{\Sigma}dS_{\Sigma}
+\phi_{\Sigma}dN_{\Sigma}
+\Phi_{\Sigma}dQ_{\Sigma}.
\end{equation}
The associated Smarr relation is
\begin{equation}
\frac{M}{2}=2T_{\Sigma}S_{\Sigma}
+\phi_{\Sigma}N_{\Sigma}
+\Phi_{\Sigma}Q_{\Sigma}.
\end{equation}
The sum sector is the charged NUT analogue of a charge sector: it contains the
charge-like variables $N$ and $Q$, but neither the ordinary angular momentum $J$ nor the
secondary hair $J_N$.
The sum sector sees $N$ directly, so the secondary hair does not replace the NUT charge.
Electric charge leaves this response visible: the potentials
$\Phi_{\Sigma}=Q/(2M)$ and $\phi_{\Sigma}=-N/M$ coexist in a simple homogeneous Smarr
relation. Thus the sum sector behaves as a system with the two charges $Q$ and $N$.

In the difference sector, the physical entropy and the sector temperature are
\begin{equation}
S_{\Delta}
=2\pi M\sqrt{M^2+N^2-Q^2-a^2},
\end{equation}
and
\begin{equation}
T_{\Delta}=
\frac{\sqrt{M^2+N^2-Q^2-a^2}}
{4\pi(2M^2+2N^2-Q^2)}.
\end{equation}

As in the uncharged construction of Eq.~(17), we now keep the remaining off-shell ambiguity explicit and write the lowest-degree homogeneous extension as
\begin{equation}
S_{\Delta}
=2\pi\sqrt{
M^4+(1+w)M^2N^2-wJ_N^2-M^2Q^2-J^2
}.
\end{equation}
The $w$-dependent term is proportional to
$M^2N^2-J_N^2$ and therefore vanishes on the physical submanifold
$J_N=MN$. Hence the physical entropy is independent of $w$, while the
off-shell derivatives need not be.

Defining
\begin{equation}
\mathcal{D}_w=
\sqrt{
M^4+(1+w)M^2N^2-wJ_N^2-M^2Q^2-J^2
},
\end{equation}
we obtain
\begin{align}
\frac{dS_{\Delta}}{2\pi}
=\frac{1}{\mathcal{D}_w}\big[
&(2M^3+(1+w)MN^2-MQ^2)dM-J\,dJ \nonumber\\
&-wJ_N\,dJ_N-M^2Q\,dQ
 +(1+w)M^2N\,dN\big].
\end{align}
On the physical family,
$\mathcal{D}_w=M\sqrt{M^2+N^2-Q^2-a^2}$, and multiplication by the
sector temperature fixed by the horizons gives
\begin{align}
T_{\Delta}dS_{\Delta}
=&\frac{1}{2(2M^2+2N^2-Q^2)}
\big[
(2M^2+(1+w)N^2-Q^2)dM \nonumber\\
&-a\,dJ-wN\,dJ_N-MQ\,dQ
 +(1+w)MN\,dN\big].
\end{align}
Since the coefficient of $dM$ must be $1/2$ in the sector first law,
the fixed horizon temperature requires
\begin{equation}
w=1
\end{equation}
for generic $N\neq0$.
Solving this identity for $dM/2$ gives the sector first law below. The explicit
derivation identifies separate $dJ$ and $dJ_N$ response terms in the difference sector
within the admissible homogeneous state space; their differentials appear in the entropy
variation with the same rotational Smarr weight.

The first law is
\begin{equation}
\frac{dM}{2}=T_{\Delta}dS_{\Delta}
+\Omega_{\Delta}dJ+\omega_{\Delta}dJ_N
+\Phi_{\Delta}\frac{dQ}{2}+\phi_{\Delta}\frac{dN}{2},
\end{equation}
with
\begin{align}
\Omega_{\Delta}&=\frac{a}{2(2M^2+2N^2-Q^2)},\\
\omega_{\Delta}&=\frac{N}{2(2M^2+2N^2-Q^2)},
\end{align}
\begin{equation}
\Phi_{\Delta}=\frac{MQ}{2M^2+2N^2-Q^2},\qquad
\phi_{\Delta}=-\frac{2MN}{2M^2+2N^2-Q^2}.
\end{equation}
The corresponding Smarr relation is
\begin{equation}
\frac{M}{2}
=2T_{\Delta}S_{\Delta}
+2\Omega_{\Delta}J+2\omega_{\Delta}J_N
+\Phi_{\Delta}\frac{Q}{2}+\phi_{\Delta}\frac{N}{2}.
\end{equation}

Electric charge appears in both sectors, as in Kerr-Newman. The difference sector also
contains the ordinary angular momentum $J$ and the NUT secondary hair $J_N$, while $N$
remains visible in the sum sector and contributes to the difference sector potential.
The charged system therefore preserves the sector evidence for the Taub-NUT bi-hair
organization within the tested class, with distinct roles for $Q$, $J$, $N$, and $J_N$.

\subsection{What reduced state spaces miss}

The full homogeneous state space closes, but that fact alone does not show that every
candidate response is needed. The reduced state spaces provide this additional test.
The metric is usually written with only $(M,a,Q,N)$, so $J_N$ could appear to be
redundant. We test this possibility directly by reexpressing the entropy in each reduced
state space before differentiating it. The comparison uses the admissible homogeneous
class defined in Sec.~\ref{sec:sector-diagnostic}.

A state space containing only $(M,Q,N)$ describes the $\Sigma$ sector but excludes
independent rotational variations. The relevant reductions therefore retain one
rotation-like channel while removing another. Define
\begin{equation}
\mathcal{H}=2M^2+2N^2-Q^2.
\end{equation}

If $J_N$ is eliminated through the physical relation $J_N=MN$ before the off-shell
differentiation, the reduced entropy in the variables $(M,Q,N,J)$ is
\begin{equation}
S_{\Delta}^{(-J_N)}
=2\pi\sqrt{M^4+M^2N^2-M^2Q^2-J^2}.
\end{equation}
On the physical family, multiplying its differential by the temperature fixed by the
horizons gives
\begin{equation}
T_{\Delta}dS_{\Delta}^{(-J_N)}
=\frac{\!(2M^2+N^2-Q^2)dM+MN\,dN-MQ\,dQ-a\,dJ\!}
{2\mathcal{H}}.
\end{equation}
The coefficient of $dM$ is
$(2M^2+N^2-Q^2)/(2\mathcal{H})$, rather than $1/2$ for generic $N\ne0$.
No potential multiplying $dN$, $dQ$, or $dJ$ can change this coefficient. Thus this
reduced embedding does not reproduce the fixed $T_{\Delta}$ first law, although it
correctly returns to Kerr-Newman when $N\to0$.

Conversely, eliminate $N$ through $N=J_N/M$ before differentiation. The reduced entropy
in $(M,Q,J,J_N)$ becomes
\begin{equation}
S_{\Delta}^{(-N)}
=2\pi\sqrt{M^4+J_N^2-M^2Q^2-J^2}.
\end{equation}
Its fixed temperature differential on the physical family is
\begin{equation}
T_{\Delta}dS_{\Delta}^{(-N)}
=\frac{(2M^2-Q^2)dM+N\,dJ_N-MQ\,dQ-a\,dJ}
{2\mathcal{H}}.
\end{equation}
Here the coefficient of $dM$ is $(2M^2-Q^2)/(2\mathcal{H})$, again different from
$1/2$ for generic $N\ne0$. This embedding also fails to reproduce the sector
temperature. Within the stated candidate class, the two calculations show that directly
eliminating either $N$ or $J_N$ does not preserve the fixed-temperature first law.

Because the entropy contains $J^2+J_N^2$, one may introduce the nonlinear composite
$\mathcal{J}=\sqrt{J^2+J_N^2}$ and rewrite the rotational part of the work one-form in
terms of $d\mathcal{J}$ on a fixed branch. This is a reparametrization of two already
identified responses, not a new elementary channel: $\mathcal{J}$ has no separate
horizon generator or known surface-integral construction, and in the enlarged state
space it replaces two variation directions by one. We therefore do not count such
nonlinear composites as candidate thermodynamic hairs. This physical restriction, rather
than the algebraic appearance of $J^2+J_N^2$, distinguishes $J$ and $J_N$ in the present
test.

Omitting $J$ while retaining $J_N$ removes the ordinary rotational response and fails to
recover the Kerr-Newman result when $N\to0$. Thus $J_N$ does not replace $J$; the charged rotating
sector contains two rotation-like response channels. Each natural reduction fails at
least one fixed temperature or homogeneous Smarr condition. This supports retaining both
NUT responses within the stated diagnostic class, without asserting uniqueness over
arbitrary off-shell completions.

\section{Sector roles and limiting checks}
\label{sec:sector-roles}

Tables~\ref{tab:sector-roles} and \ref{tab:limits} summarize the sector assignments and
their limiting regimes. Charge-like variables appear in the $\Sigma$ sector, whereas
closure of the $\Delta$ sector supports the rotation-like variables. Because the charged
state space already contains the gauge charge $Q$ and angular momentum $J$, it tests
whether the two NUT responses remain distinct in their presence.

\begin{table*}[t]
\begin{minipage}[t]{0.47\textwidth}
\caption{\label{tab:sector-roles}
Thermodynamic roles supported by the Kerr-Newman-NUT sector first laws. The weights
refer to length scaling in the homogeneous equation of state.}
\centering
\small
\begin{ruledtabular}
\begin{tabular}{ccc}
variable & weight & sector role \\
$Q$ & one & charge-like, both sectors \\
$N$ & one & NUT charge-like; $\Sigma$ closure \\
$J$ & two & ordinary rotation, $\Delta$ sector \\
$J_N$ & two & NUT secondary hair, $\Delta$ sector
\end{tabular}
\end{ruledtabular}
\end{minipage}
\hfill
\begin{minipage}[t]{0.47\textwidth}

\caption{\label{tab:limits}
Limiting regimes of the sector thermodynamic state space. The listed variables are those
retained by the sector first law and Smarr closure tests for the family considered here, not
independent metric parameters.}
\centering
\small
\begin{ruledtabular}
\begin{tabular}{cc}
limit & resulting sector structure \\
Kerr-Newman-NUT & $(Q,N,J,J_N)$ \\
$N\to0$ & Kerr-Newman $(Q,J)$ \\
$Q\to0$ & Kerr-NUT $(N,J,J_N)$ \\
$Q,a\to0$ & Taub-NUT $(N,J_N)$ \\
$N,Q\to0$ & Kerr $(J)$ \\
$N,a\to0$ & Reissner-Nordstr\"om $(Q)$
\end{tabular}
\end{ruledtabular}
\end{minipage}
\end{table*}

There are two distinct meanings of ``appears in a sector'' in
Table \ref{tab:sector-roles}. A variable can
enter the sector entropy itself, changing the sector equation of state, or it can enter
through its conjugate potential in the sector first law. The electric charge $Q$ does
both: it contributes to $S_{\Sigma}$ and $S_{\Delta}$ and has potentials in both sector
first laws. The NUT charge $N$ also contributes to both entropies, but its most direct
charge-like role is exposed by the $\Sigma$ sector, where the first law closes with
$N_{\Sigma}=N/2$ and no rotation-like NUT hair. The variables $J$ and $J_N$, by
contrast, are absent from $S_{\Sigma}$ and from the $\Sigma$ first law. They enter only
through the difference sector equation of state and its Smarr relation, which defines
the rotation-like role of $J_N$.

$J_N$ cannot be identified with either $Q$ or $J$. The electric charge appears with
potentials $\Phi_{\Sigma}$ and $\Phi_{\Delta}$ and has
weight one in the sector Smarr relations. The ordinary angular momentum appears with
$2\Omega_{\Delta}J$ and has weight two. The secondary NUT hair has the same Smarr weight
as $J$, but its conjugate potential is $\omega_{\Delta}$ rather than
$\Omega_{\Delta}$, and it is tied to the NUT charge through the physical submanifold
$J_N=MN$. It is rotation-like in scaling but NUT-like in origin.

For variations restricted to the charge-like variables $(Q,N)$, the response is
captured already by the $\Sigma$ sector. If the variation probes the horizon separation
information encoded in the entropy difference, then the variables with angular momentum scaling
$(J,J_N)$ enter as distinct response channels in the tested representation. The two roles
are not redundant within this class because they are tested by
different sector temperatures: $T_{\Sigma}=1/(8\pi M)$ and
$T_{\Delta}=\sqrt{M^2+N^2-Q^2-a^2}/[4\pi(2M^2+2N^2-Q^2)]$. A redefinition that hides
$J_N$ inside $N$ would have to reproduce both temperatures and both Smarr formulae
simultaneously; the direct reduced embedding tests above do not achieve this within the
restricted homogeneous state space.

The limiting regimes listed in Table \ref{tab:limits} separate two statements that are
sometimes conflated. The first is continuity:
the formulae reduce to the known Kerr-Newman and Taub-NUT sectors in the corresponding
limits. The second concerns the tested reductions: in the electrically charged
Kerr-Newman-NUT family considered here, omitting the explicitly tested response channel
prevents at least one sector first law or Smarr relation from closing.

The limiting cases check the formulae. When $N\to0$, the
secondary hair and its potential vanish, while
\begin{align}
S_{\Delta}&\to2\pi\sqrt{M^4-M^2Q^2-J^2},\\
T_{\Delta}&\to\frac{\sqrt{M^2-Q^2-a^2}}{4\pi(2M^2-Q^2)},
\end{align}
and the Kerr-Newman difference sector first law of Sec.~\ref{sec:benchmarks} is recovered. When
$Q,a\to0$, the difference sector entropy reduces to
$2\pi M\sqrt{M^2+N^2}$ and the potentials reduce to the Taub-NUT prototype values in
Sec.~\ref{sec:tn-prototype}. If $Q\to0$ but $a$ remains nonzero, the system keeps both
rotation-like variables $J$ and $J_N$, consistently treating ordinary angular momentum
and the NUT secondary hair as distinct response entries in the candidate sector state
space even though both are rotation-like in their Smarr weights.

The sum sector has analogous limits. For $N\to0$,
\begin{equation}
S_{\Sigma}\to\pi(2M^2-Q^2),\qquad
T_{\Sigma}\to\frac{1}{8\pi M},\qquad
\Phi_{\Sigma}\to\frac{Q}{2M},
\end{equation}
which is exactly the Kerr-Newman sum sector. For $Q\to0$,
\begin{equation}
S_{\Sigma}\to2\pi(M^2+N^2),\qquad
\phi_{\Sigma}\to-\frac{N}{M},
\end{equation}
which is the charge-like Taub-NUT sector. These reductions show that the charged NUT
sector thermodynamics interpolates between two independently understood structures
rather than replacing either of them.

The Kerr-NUT limit $Q\to0$ but $a\ne0$ is also informative. The sum sector becomes
\begin{equation}
S_{\Sigma}\to2\pi(M^2+N^2),\qquad
T_{\Sigma}\to\frac{1}{8\pi M},
\end{equation}
while the difference sector retains both variables with angular momentum scaling,
\begin{equation}
S_{\Delta}\to
2\pi\sqrt{M^4+2M^2N^2-J_N^2-J^2}.
\end{equation}
Kerr-NUT does not collapse to either the Kerr benchmark or the Taub-NUT prototype:
it contains both the ordinary rotational hair $J$ and the NUT secondary hair $J_N$,
making their distinct roles explicit.

The Reissner-Nordstr\"om limit $N,a\to0$ gives the opposite check. In this case
$J=J_N=0$, the difference sector contains no rotation-like variable, and the remaining
sector thermodynamics is controlled only by $M$ and $Q$. In the Schwarzschild
limit $Q,N,a\to0$, the sum and difference constructions become degenerate in the sense
that no additional charge or rotation variable remains to be diagnosed. These limits
confirm that the adopted sector representation reduces continuously and retains no
rotation-like variable when the corresponding geometric parameters vanish.

The proposed state space would fail the test if the
$\Delta$ sector could be written with the fixed temperature $T_{\Delta}$ and a valid
Smarr relation using only $(M,Q,N,J)$; the secondary hair would then be unnecessary. If
it could be written using $(M,Q,J,J_N)$ while dropping the charge-like NUT variable, the
bi-hair interpretation would also fail. For the natural reduced embeddings tested above,
neither omission reproduces the first law coefficient fixed by $T_{\Delta}$; the
associated homogeneous scaling test fails as well. This supports retaining both $N$ and
$J_N$ within the stated diagnostic class; it does not exclude other homogeneous
off-shell completions.
The role of $J_N$ follows primarily from matching the first law coefficient to the
temperature obtained from the horizon harmonic mean. The Bekenstein-Smarr relation then
checks the assigned scaling weights. The same homogeneous state space has the correct
Kerr-Newman and
Taub-NUT limits. The secondary response therefore remains supported after adding
ordinary electric charge and rotation.

\section{Rotating parent hidden CFT normalization}
\label{sec:hidden-cft}

We compare the sector temperatures with those obtained from the hidden conformal
symmetry of the rotating Kerr-Newman-NUT parent family
\cite{PRD82-066004,JHEP080087,PRD82-124051,JHEP1010074,PTP125-47,NPB848-108,MPLA27-1250046,
IJMPD27-1850109,NPB953-114970,IJMPA35-2050156,PLB827-136892,JHEP1024078,JHEP1024089}.
The hidden conformal construction organizes the near region scalar wave equation in
terms of two thermal CFT sectors and depends on the same horizon combinations that enter
the sum and difference thermodynamics. This comparison checks the normalization of the
sector split; it does not derive $J_N$ microscopically or assume a dual CFT.

We use the rotation, or $J$-picture, of the hidden conformal construction. Other charge
pictures can reorganize the microscopic labels, but they are not needed for the sector
thermodynamic criterion used here. In this $J$-picture, the microscopic temperatures
for the charged Kerr-Newman-NUT geometry may be written as
\begin{align}
\mathcal{T}_R&=\frac{\sqrt{M^2+N^2-Q^2-a^2}}{2\pi a},\\
\mathcal{T}_L&=\frac{2M^2+2N^2-Q^2}{4\pi Ma}.
\end{align}
Here the labels $R$ and $L$ follow the standard hidden conformal convention. In the
present thermodynamic notation the corresponding horizon combinations are the
difference and sum sectors, respectively: the square root in $\mathcal{T}_R$
is the same horizon separation factor that enters $T_{\Delta}$, whereas
$\mathcal{T}_L$ contains the same symmetric combination
$2M^2+2N^2-Q^2$ that enters the $\Sigma$ sector equation of state. The two sets of
temperatures are not expected to be identical, because the hidden conformal
temperatures are conjugate to dimensionless microscopic CFT variables whose
normalization depends on the rotation parameter $a$. The thermodynamic sector
temperatures, by contrast, are
defined directly from the horizon first laws with sector energy $M/2$.

With this normalization difference taken into account, the two ratios agree:
\begin{equation}
\frac{\mathcal{T}_R}{T_{\Delta}}=
\frac{\mathcal{T}_L}{T_{\Sigma}}
=\mathcal{R}
=\frac{2(2M^2+2N^2-Q^2)}{a}.
\end{equation}
At fixed rotating background, the common factor $\mathcal{R}$ relates the
hidden conformal pair to the
thermodynamic sector pair,
\begin{equation}
\mathcal{T}_L=\mathcal{R}T_{\Sigma},\qquad
\mathcal{T}_R=\mathcal{R}T_{\Delta}.
\end{equation}
The left-moving hidden conformal temperature aligns with the sum sector, and the
right-moving temperature with the difference sector. Kerr and Kerr-Newman have the same
pattern: the symmetric horizon combination describes the nonrotating or charge-like
sector, while the antisymmetric combination detects rotation.

The charged NUT parameters enter the same normalization check. The electric charge enters
both hidden conformal temperatures through the combination $Q^2$, just as it enters both
$S_{\Sigma}$ and $S_{\Delta}$. The NUT charge enters through the replacement
$M^2\to M^2+N^2$ in the symmetric combinations, while the difference sector still
retains the rotation-like dependence through $a$. This supports the relative
$\Sigma/\Delta$ normalization obtained from the first law analysis. It does not test
whether $J_N$ must be retained as an independent thermodynamic response channel; that
question is addressed only by the sector closure calculation above.

Because the conventional hidden conformal temperatures contain $1/a$, their strictly
nonrotating Taub-NUT limit is singular. The nonrotating state space is therefore derived
from the two-horizon thermodynamics, whose sector temperatures and first laws have a
regular limit. For $a\neq0$, hidden conformal symmetry supplies only a normalization
check: its microscopic temperature pair has the same relative organization as the
thermodynamic sectors.

\section{Discussion}
\label{sec:discussion}

The charged Kerr-Newman-NUT family retains the sector pattern of the uncharged Taub-NUT
case. Within the physically anchored minimal homogeneous class used here, the $\Sigma$
sector is governed by the charge-like variables $Q$ and $N$, while the $\Delta$ sector
also requires the rotation-like variables $J$ and $J_N$. This supports the same bi-hair
organization after the standard Kerr-Newman variables are included.
The horizon temperatures fix the sector temperatures before the thermodynamic potentials
are chosen. Matching the corresponding first law coefficient is the main state-space
test, while the Bekenstein-Smarr relation checks homogeneous scaling. The natural
reductions examined above cannot absorb the NUT secondary response into electric charge
or ordinary angular momentum.
These failures connect the closure test to the interpretation of the result: directly
eliminating either NUT response does not preserve the temperature coefficient fixed by
the horizons, and the corresponding homogeneous Smarr relation also fails.

Within this diagnostic, $J_N=MN$ is a thermodynamic secondary hair in the homogeneous
equation of state. It is neither a new metric parameter nor an independent asymptotic
Noether charge. The physical Kerr-Newman-NUT family lies on the submanifold $J_N=MN$,
while derivatives with respect to $J_N$ probe the enlarged state space. A charge
interpretation based on subleading dual charges is not assumed here. Nevertheless, the
background-subtracted Komar-type construction~\cite{PRD108-064035} gives $J_N$
a surface-integral origin and motivates treating it as a physically identified candidate
response rather than an arbitrary nonlinear combination of thermodynamic variables. Its relation to asymptotic
subleading dual charges \cite{JHEP0119143,JHEP0319057} requires further study.
This conclusion concerns the two-horizon state-space question and should be distinguished
from the choice of a complete macroscopic formulation of NUT thermodynamics.

Other formulations of NUT thermodynamics organize different macroscopic first laws.
They may emphasize the Misner string, choose a different thermodynamic energy, or use
homogeneous response channels. The sector diagnostic instead asks which variables close
the thermodynamics once
$S_{\Sigma}$, $S_{\Delta}$, $T_{\Sigma}$, and $T_{\Delta}$ have been fixed by the
two horizons. The appearance of $J_N$ answers this question without rejecting other
macroscopic formulations. It does not determine the full Misner string first law, settle the global charge
interpretation of the NUT parameter, or choose a unique macroscopic ensemble.

The limits $N\to0$ and $Q,a\to0$ recover the Kerr-Newman and Taub-NUT results,
respectively. The Christodoulou-Ruffini-type mass formula and entropy product involve the
same homogeneous variables as the horizon and sector first laws. In the rotating
$J$-picture, the hidden conformal temperatures also reproduce the relative normalization
of the $\Sigma$ and $\Delta$ sectors, although their nonrotating limit is singular.

Natural extensions include dyonic, AdS, higher dimensional, and more general NUT geometries.
A dyonic extension would test whether electric and magnetic charges enter
the two sectors symmetrically, while an AdS extension would bring in pressure-volume
terms and possible string contributions; hidden conformal and warped-CFT descriptions
of NUT-AdS and Pleba\'nski-Demia\'nski families are useful comparison points
\cite{NPB848-108,PTP125-47,NPB953-114970,JHEP1024089}. Higher dimensional NUT
solutions would test whether multiple NUT parameters generate a corresponding set of
secondary hairs. In each case one would apply the same diagnostic: determine the
sector temperatures fixed by the horizons, choose an admissible homogeneous state space, and
test simultaneous closure of the sector first laws and Smarr relations.

\begin{acknowledgments}
This work is supported by the National Natural Science Foundation of China (NSFC) under
Grants No. 12675077, No. 12205243, and No. 12375053, by China Scholarship Council (CSC), by the Sichuan
Science and Technology Program under Grant No. 2026NSFSC0021, and by the Natural
Sciences and Engineering Research Council of Canada.
\end{acknowledgments}

\end{document}